\documentclass{elsart}

\usepackage{amsmath}

\begin{document}

\begin{frontmatter}

\title{Crossover from thermal to quantum creep in layered antiferromagnetic superconductor}

\author{Tomasz Krzyszto\'{n}.}

\address{Institute for Low Temperatures and Structure Research, Polish Academy of
Sciences, 50-950 Wroc{\l }aw, Poland.}
\begin{abstract}
The influence of the antiferromagnetic order on the superconductor in the
mixed state results in creation of spin-flop domains along the cores of the
vortex lines. It is shown that this phenomenon makes possible crossover from
quantum creep regime to thermal one, and vice versa, at constant
temperature. To do this one needs to simply change the intensity or the
direction of applied magnetic field in the basal ${\bf ab}$ plane of layered
structure.
\end{abstract}


\end{frontmatter}

\section{Introduction}

The suppression of superconductivity by applied magnetic fields implies that
magnetism and superconductivity are two very different cooperative phenomena
that are unlikely to be present simultaneously within the same sample. On
the other hand, there exist materials in which both competitive phenomena do
coexist in the same volume. This happens because the 4f electrons of rare
earth ($RE$) ions, in most cases responsible for magnetism, and those
electrons responsible for superconductivity are spatially well-separated in
regular crystal lattice\cite{1,2}. The specific feature caused by the
long-range antiferromagnetic order in the mixed state of superconductor is
the creation of spin-flop (or metamagnetic) domain along each vortex core %
\cite{3}-\cite{6}. This is easy to understand by taking two sublattices
antiferromagnet as an example. An infinitesimal magnetic field applied
perpendicular to the easy axis makes the ground antiferromagnetic state
unstable against the phase transformation to the canted phase (spin-flop).
On the contrary, if the magnetic field is applied parallel to the easy axis
the antiferromagnetic configuration is stable up to the thermodynamic
critical field $H_{T}$. When the field is further increased a canted phase
develops in the system. Assume that in the antiferromagnetic superconductor
the lower critical field fulfils the relation $H_{c1}<%
{\frac12}%
H_{T}$ and that the external field,$H_{c1}<H<%
{\frac12}%
H_{T}$ , is applied parallel to the easy axis. Then the superconducting
vortices appear in the ground antiferromagnetic state. What happens if the
field is increased above $%
{\frac12}%
H_{T}$? The phase transition to the canted phase originates in the vortex
core because the field intensity in the core doubles the external one. The
spatial distribution of the field around the vortex is a decreasing function
of the distance from its center. Hence the magnetic field intensity in the
neighborhood of the core is less then $H_{T}$. Therefore, the rest of the
vortex remains in the antiferromagnetic configuration. The radius of spin
flop domain grows as the field is increased. There exists critical field for
flux penetration in the form of vortices with canted phase along the core %
\cite{3,4}. When the external field is applied perpendicular to the easy
axis, the vortices do not change their structure and behave quite similarly
to the ones in nonmagnetic superconductor. Thus, in the considered model
there are two distinct types of vortices. The above considerations apply to
classical superconducting Chevrel phases as well as to the high $T_{c}$
superconductors, where antiferromagnetic order is produced by the regular
lattice of rare earth ($RE$) ions occupying isolating layers. A typical
example of such system is $ErBa_{2}Cu_{3}O_{7}$. This compound has
tetragonal unit cell with small orthorombic distortion in the $ab$ plane.
The $Er$ ions form two sublattices antiferromagnetic structure of magnetic
moments laying parallel and antiparallel to the $\mathbf{b}$ direction \cite%
{7}. Another example may be $RE$ nickel boride- carbides \cite{8}. Their
structure consists of $RE$-carbon layers separated by $Ni_{2}B_{2}$ sheets.
The theoretical structure used in the calculations in the present model is
shown in Fig.1. It consists of superconducting layers of thickness $d_{s}$
and isolating ones of thickness $d_{i}$, $d=d_{i}+d_{s}$. In the isolating
layers, the magnetic moments are running parallel and antiparallel to the $%
\mathbf{b}$ direction (easy axis). The magnetic field aligned parallel to
the conducting planes makes the vortex lattice to accommodate itself to the
layer structure so that the vortex cores lie in between the superconducting
sheets. A current density $j$, flowing along the planes perpendicular to the
applied magnetic field exerts a Lorentz force on the vortices in the $%
\mathbf{c}$ direction so that intrinsic pinning barriers are formed on
strongly superconducting layers. This paper is devoted to the problem of
resistive properties of the system caused by the quantum creep. In the
previous paper \cite{6} thermally assisted flux motion was considered. The
experimental evidence of quantum tunneling is based on the fact that the
magnetic moment relaxation rate exhibits two types of behavior as a function
of temperature. Above a characteristic temperature $T_{0}$ in the thermal
activation regime the decay rate is of the Arrhenius type $\Gamma \sim \exp
\left( -U_{0}/k_{B}T\right) $. Below $T_{0}$, the decay rate is essentially
independent of temperature$\Gamma \sim \exp \left( -S/\hbar \right) $ and is
interpreted as arising from the quantum tunneling of vortices through
intrinsic pinning potential \cite{9,10}. The purpose of the present
calculations is to show drastic change of tunneling rate and crossover
temperature due to the phase transition to the canted phase around the
vortex core.

\section{CALCULATION OF CROSSOVER TEMPERATURE}

Consider the vortex line as a straight string-like object of an effective
mass $M$ per unit length trapped into a metastable state in an intrinsic
pinning potential $V(u)$ and exposed to continuous deformation $u(x,t)$ in
the $\mathbf{z}$ direction. The magnetic field is applied in $\mathbf{x}$
direction ($\mathbf{b}$ direction on Fig.1). In the semi-classical
approximation the quantum decay rate is calculated as a saddle-point
solution (bounce) of the Euclidean action $S$ for the string 
\begin{eqnarray}
S=\int_{-\infty }^{\infty }dx\int_{0}^{\hbar \beta }\left\{ d\tau 
{\frac12}%
M\left( \frac{\partial u}{\partial \tau }\right) ^{2}+\frac{\varepsilon _{l}%
}{2}\left( \frac{\partial u}{\partial x}\right) ^{2}+V(u)\right\} \nonumber \\ 
-\int_{-\infty }^{\infty }dx\int_{0}^{\hbar \beta }\left\{ \frac{\eta }{2\pi }%
\frac{\partial u}{\partial \tau }\int_{0}^{\hbar \beta }d\tau ^{^{\prime }}%
\frac{\partial u}{\partial \tau ^{^{\prime }}}\ln \left| \sin \frac{\pi }{%
\hbar \beta }\left( \tau -\tau ^{^{\prime }}\right) \right| \right\}. 
\end{eqnarray}%
Here $\beta =\left( k_{B}T\right) ^{-1}$ , $\varepsilon _{l}$ is the line
tension of the vortex, $\eta $ is the viscosity coefficient and $\tau $
denotes imaginary time. The pinning potential $V(u)$ consists of intrinsic
periodic part and the Lorentz potential: 
\begin{equation*}
V\left( u\right) =-\frac{\varphi _{0}j_{c}d}{2\pi }\cos \left( \frac{2\pi u}{%
d}\right) -\varphi _{0}ju
 .
\end{equation*}%
For large current, this potential can be expanded around the inflection
point to give 
\begin{equation}
V\left( u\right) =V_{0}\left[ \left( \frac{u}{w}\right) ^{2}-\left( \frac{u}{%
w}\right) ^{3}\right], 
\end{equation}%
where $V_{0}=\frac{2}{3}\frac{\varphi _{0}j_{c}^{2}\pi ^{2}}{d^{2}}w^{3}$and 
$w=\frac{3d}{\pi }\left( \frac{j_{c}-j}{2j_{c}}\right) ^{\frac{1}{2}}$may be
thought as the width of the barrier because $V(0)=V(w)=0$, $j_{c}$ is the
critical depinning current. The last term in Eq.(1) is the so-called
Caldeira-Leggett action \cite{11}, which describes ohmic damping produced by
the coupling to the heat-bath of harmonic oscillators. The line tension$\
\varepsilon _{l}$ is different for vortices in two different orientations in
the $\mathbf{ab}$ plane. As discussed above the vortices lying parallel to $%
\mathbf{a}$ direction and those laying in the $\mathbf{b}$ direction but
created in the magnetic field fulfilling relation $H_{c1}<H<%
{\frac12}%
H_{T}$ \ \cite{6} have the line tension equal to 
\begin{equation}
\varepsilon _{l}=\varepsilon _{a}=\varepsilon _{0}\ln \frac{\lambda _{ab}}{d}
,
\end{equation}%
 where \ $\varepsilon _{0}=\frac{\varphi _{0}^{2}}{16\pi ^{2}\lambda
_{ab}^{2}\mu _{0}}$ . For those vortices lying in the $\mathbf{b}$ direction
but possessing spin flop domain, we write the following expression \cite{5} 
\begin{equation}
\varepsilon _{l}=\varepsilon _{b}=\frac{\varphi _{0}H_{T}}{2}+\frac{9}{128}%
\varepsilon _{0}\ln \frac{\kappa ^{2}\varphi _{0}}{\pi \lambda _{j}^{2}B_{T}}
.
\end{equation}%
Here $B_{T}=H_{T}+2M_{0}$, $2M_{0}$ is the magnetization of the spin flop
domain, $\lambda _{j}=\frac{\lambda _{c}}{\lambda _{ab}}d\ \ $, $\lambda $
denotes penetration depth of the magnetic field and $\kappa $ is the
Ginzburg-Landau parameter in the $\mathbf{ab}$ plane. In the semiclassical
approximation the decay rate is given by the value of the action on a
classical trajectory obtained from the Euler-Lagrange equations of the
motion 
\begin{equation}
-M\frac{\partial ^{2}u}{\partial \tau ^{2}}-\varepsilon _{l}\frac{\partial
^{2}u}{\partial x^{2}}+V^{^{\prime }}(u)+\frac{\eta }{\hbar \beta }%
\int_{0}^{\hbar \beta }d\tau \frac{\partial u}{\partial \tau }\cot \frac{\pi 
}{\hbar \beta }\left( \tau -\tau ^{^{\prime }}\right) =0
 .
\end{equation}%
The trajectory $u_{0}(x)$ for static solution of Eq.(5) gives the
activation energy in the thermal regime $T>T_{0}$. Below this crossover
temperature a new kind of trajectory, periodic in imaginary time, develops.
Therefore, $u(x,\tau )$ can be expanded in the Fourier series with Matsubara
frequencies 
\begin{equation}
u(x,\tau )=\sum_{n=0}^{\infty }u_{n}\left( x\right) \cos \left( \omega
_{n}\tau \right) \ \ \ ;\ \ \ \omega _{n}=\frac{2\pi n}{\hbar \beta }.
\end{equation}%
Substituting this expansion into Eq.(5) and linearizing potential around the
static solution $u_{0}(x)$ one obtains 
\begin{equation}
-\varepsilon _{l}\frac{\partial ^{2}u_{n}}{\partial x^{2}}+V^{^{\prime
\prime }}(u_{0})u_{n}=-\left( \eta \omega _{n}-M\omega _{n}^{2}\right) u_{n}.
\end{equation}%
Upon introducing new variables $v_{n}=\frac{u_{n}}{w}$ \ and \ $\zeta =\frac{%
x}{d}\left( \frac{\pi ^{2}w\varphi _{0}j_{c}}{\varepsilon _{l}}\right) ^{%
\frac{1}{2}}$ the static equation now reads. 
\begin{equation*}
-\frac{1}{2}\frac{\partial ^{2}v_{0}}{\partial \zeta ^{2}}%
+2v_{0}-3v_{0}^{2}=0
 .
\end{equation*}%
Its solution is easily found to be 
\begin{equation}
v_{0}=\cosh ^{-2}\zeta 
.
\end{equation}%
Substitution Eq.(8) into Eq.(7) results in the following equation 
\begin{equation}
-\frac{1}{2}\frac{\partial ^{2}v_{n}}{\partial \zeta ^{2}}+2\left( 1-3\cosh
^{-2}\zeta \right) v_{n}=E_{n}v_{n},
\end{equation}%
where 
\begin{equation}
E_{n}=-\frac{j_{c}w^{2}}{V_{0}^{2}}\left( \eta \omega _{n}+M\omega
_{n}^{2}\right) .
\end{equation}%
Eq.(10) has three discrete eigenvalues $-5/2,0,3/2$ \cite{12}. The negative
one determines the crossover temperature 
\begin{equation}
k_{B}T_{0}=\frac{\hbar \eta }{4\pi M}\left\{ \left[ 1+\frac{20\pi \varphi
_{0}j_{c}M}{d\eta ^{2}}\left( \frac{j_{c}-j}{2j_{c}}\right) ^{\frac{1}{2}}%
\right] ^{\frac{1}{2}}-1\right\} 
\end{equation}

\section{ESTIMATION OF EFFECTIVE MASS AND VISCOSITY}

The above calculations apply to both kinds of vortices. The only difference
is their effective mass and viscosity coefficient. It is possible to express
these parameters as the function of condensation energy accumulated in the
vortex cores. For the stationary flux flow at $j>j_{c}$ the viscous force $%
\eta \frac{\partial u}{\partial t}$\ is equal to Lorentz force. The electric
field generated by the moving vortex is $E=B\frac{\partial u}{\partial t}$,
so we get $E=\frac{\varphi _{0}B}{\eta }j=\rho j=\rho _{N}\frac{B}{H_{c2}}j$%
\ where $\rho _{N}$ is the normal phase resistivity in the $ab$ plane and $%
H_{c2}$ is the upper critical field parallel to the layers. Finally, 
\begin{equation}
\eta =\frac{\varphi _{0}H_{c2}}{\rho _{N}}=\frac{\varphi _{0}\kappa H_{c}%
\sqrt{2}}{\rho _{N}}=\varepsilon _{l}\frac{4\sqrt{3}\kappa ^{2}}{\pi \rho
_{N}\ln \kappa },
\end{equation}
where $H_{c}=\frac{\varepsilon _{l}\kappa 2\sqrt{6}}{\pi \varphi _{0}\ln
\kappa }$\ \ is calculated from the constitutive relation $\varepsilon
_{l}=H_{c1}\varphi _{0}$. The effective mass of the vortex can be deduced
from the work of Suhl \cite{13}. He derived the core contribution to the
inertial mass $m_{core}=\frac{3}{8}m_{e}\frac{\xi ^{2}H_{c}^{2}\mu _{0}}{%
\epsilon _{F}}$, where $m_{e}$ denotes the mass of the electron and $%
\epsilon _{F}$ is the Fermi energy, and the electromagnetic contribution
coming from the energy of the electric field induced by the moving flux.
Simple estimation shows that this contribution in layered superconductors is 
$10^{-4}$ of the core contribution. Therefore, 
\begin{equation}
M=\varepsilon _{l}^{2}\frac{9\lambda _{ab}^{2}m_{e}\mu _{0}}{\varphi
_{0}^{2}\pi ^{2}\epsilon _{F}\left( \ln \kappa \right) ^{2}}.
\end{equation}
It is possible now to relate the crossover temperature in Eq.(11) to the
line tension of the vortex 
\begin{equation}
T_{0}=\alpha \varepsilon _{l}^{-1}.
\end{equation}
The coefficient $\alpha $ depends on the material constants and current
intensity $j.$

\section{CROSSOVER FROM QUANTUM TO THERMAL CREEP}

As was mentioned in the introduction there are two tapes of vortex lines in
the system. The first ones, without magnetic domain, occur when the field is
applied in the $\mathbf{a}$ or $\mathbf{b}$ directions, but its intensity
does not exceed $%
{\frac12}%
H_{T}$. Eq.(3) gives their line tension and the related crossover
temperature is given by $T_{0a}=\alpha \varepsilon _{a}^{-1}$. The other
type, possessing magnetic domain, occurs when the field is applied in the $%
\mathbf{b}$ direction and its intensity exceeds $%
{\frac12}%
H_{T}$. Eq.(4) gives their line tension and the crossover temperature is $%
T_{0b}=\alpha \varepsilon _{b}^{-1}$. It is easy to see that $\varepsilon
_{b}>\varepsilon _{a}$ and therefore $T_{0a}>T_{0b}$ . The above
calculations lead to the following conclusion. It is possible to switch the
creep regime at constant temperature. To do this, one needs to change the
field intensity or simply change the field direction in the $\mathbf{ab}$
plane. The diagrams in Fig.2 show possible scenarios of crossover from
quantum to thermal regimes. Let us discuss just two of these scenarios.
Here,the first prescription is as follows. Fix the temperature $T_{0}$
somewhere in the range $T_{0a}>T_{0}>T_{0b}$. Then align the external field
in the $\mathbf{a}$ direction and increase its intensity to the point marked
II on the upper diagram in Fig.2. The system is in the quantum creep regime
now. Then move the direction of external field from $\mathbf{a}$ to $\mathbf{%
b}$ axis. The system jumps to the point II of the lower diagram and finds
itself in the thermal creep regime. Doing the same operations in the reverse
order one enforces the system to crossover from thermal to quantum creep
regime. The other scenario is the following. Apply magnetic field along $%
\mathbf{b}$ axis and increase its intensity above $%
{\frac12}%
H_{T}$ keeping temperature constant in the interval. $T_{0a}>T_{0}>T_{0b}$.
Now, the system switches from quantum to thermal regime.In other words, the
system moves along the line defined by points I and II on the lower diagram
of Fig.2.

\section{ACKNOWLEDGEMENTS}

This work was supported by Komitet Bada\'{n} Naukowych under grant 2 PO3B
125 19

FIGURE CAPTIONS

Fig.1. Schematic drawing of a layered antiferromagnetic superconductor. The
rare earth elements form two sublattices of magnetic moments (bold arrows)
lying parallel and antiparallel to b direction. The reference frame and the
crystallographic axes are shown.

Fig.2. Schematic diagrams showing possible ways ( marked by arrows ) of
changing quantum creep behavior in the system to thermal one, and vice
versa. Shaded areas on the diagrams correspond to quantum creep regime.\-

\end{document}